\documentclass[aps,prl,twocolumn,showpacs]{revtex4}

\usepackage{amsmath}
\usepackage{amssymb}
\usepackage{graphics}

\begin{document}


\title{Universality in the entanglement structure of ferromagnets}

\author{J.S. Pratt}

\affiliation{
Center for Quantum Information and
Department of Physics and Astronomy,
University of Rochester, Rochester, NY 14627}

\begin{abstract}
    Systems of exchange-coupled spins are commonly used to model
    ferromagnets.  The quantum correlations in such magnets are
    studied using tools from quantum information theory.  Isotropic
    ferromagnets are shown to possess a universal low-temperature
    density matrix which precludes entanglement between spins, and the
    mechanism of entanglement cancellation is investigated, revealing
    a core of states resistant to pairwise entanglement cancellation. 
    Numerical studies of one-, two-, and three-dimensional lattices as
    well as irregular geometries showed no entanglement in
    ferromagnets at any temperature or magnetic field strength.
\end{abstract}


\maketitle


Quantum correlations are responsible for much of the challenge in
understanding interacting many-body quantum systems, and it is
therefore of fundamental importance to have quantitative knowledge of
these correlations.  Progress in quantum information theory has led to
the development of new measures of the inseparability of a quantum
state, and in the last few years these measures have been used to
assess the quantum correlations in diverse physical systems. 
Concurrence \cite{concurrence} is an especially useful metric for such
studies because it can be applied to mixed as well as pure states.  It
therefore can be used to quantitate the thermal entanglement in a
system at nonzero temperature.  It can also be applied to evaluate the
inseparablilty of an equal incoherent mixture of degenerate energy
eigenstates.  However, concurrence is defined only for a pair of
qubits.  Since a qubit is formally equivalent to a spin-$1/2$ particle
when only the spin degree of freedom of the latter is considered, this
has led to several analyses of the thermal entanglement between a pair
of interacting spin-$1/2$ particles.  Two-spin systems studied in this
way include the $XXX$ \cite{ArnesenBoseVedral}, $XXZ$ \cite{Wang1}, and
$XYZ$ \cite{ZhouSongGuoLi} Heisenberg models, the $XX$-model \cite{Yeo},
the $XY$-model \cite{KamtaStarace}, the Heisenberg-Dzyaloshinski-Moriya
model \cite{Wang1}, and the Ising model \cite{GunlyckeKendonVedralBose}. 
Inseparability at zero and finite temperture has also been studied in
two-spin subsystems of one-dimensional spin chains, including the
$XXX$-chain \cite{ArnesenBoseVedral,Wang2,LatorreRicoVidal}, the
$XY$-chain \cite{Wang3,OsterlohAmicoFalciFazio,OsborneNielsen}, the
Ising chain \cite{GunlyckeKendonVedralBose}, and the Majumdar-Ghosh
chain \cite{BoseChattopadhyay}, as well as in the ground state of
quasi-one-dimensional spin ladders \cite{BoseChattopadhyay}.

This paper will examine quantum correlations in isotropic ferromagnets
without restrictions on the geometry of the spins (such as
one-dimensionality or periodicity) or on the range of the spin-spin
interaction.  A suitably general model of a magnet, which will be
called the Heisenberg spin graph, consists of $N$ spins coupled by
exchange interactions with arbitrary range and strength.  It is
defined by the Hamiltonian

\begin{equation}\label{HSGHamiltonian}
    H_{HSG}=\sum_{i<j} J_{ij} {\mathbf{S}}_{i} \boldsymbol{\cdot}
    {\mathbf{S}}_{j} .
\end{equation}
This can be visualized as a graph with a spin-$1/2$ particle at each
vertex, where the edge between spins $i$ and $j$, representing the
exchange interaction ${\mathbf{S}}_{i} \boldsymbol{\cdot}
{\mathbf{S}}_{j}$ between the two spins, is weighted by the coupling
constant $J_{ij}$.  The summation runs over each pair of spins once. 
The operator ${\mathbf{S}}_{i}=(S_{i}^{x},S_{i}^{y},S_{i}^{z})$ is the
spin operator associated with the spin at the $i$th vertex of the
graph; its Cartesian components obey the usual angular momentum
algebra.  As a matter of notation, states in the $2^{N}$-dimensional
Hilbert space ${\mathcal{H}}$ of the model are specified with respect
to basis states of the form $\vert i,j, \ldots, k \rangle$, where the
listed spins $i,j,\ldots,k$ are aligned in the $+z$-direction (`up')
and the remaining spins are antiparallel (`down').  $\vert \emptyset
\rangle$ denotes the state with all spins down.  One might anticipate
that the Heisenberg spin graph would be so general a model that
nothing much could be said about it.  It turns out that when the
coupling constants $J_{ij}$ are nonpositive (that is, ferromagnetic),
certain features of Heisenberg spin graphs are universal, in the sense
that they do not depend on the exact values of the coupling constants
or even on the number of spins.  This universality allows a complete
analysis of low-temperature correlations in such ferromagnets.

Because the Heisenberg spin graph Hamiltonian $H_{HSG}$ commutes with
the total $z$-spin operator $S^{z} = \sum_{i=1}^{N} S_{i}^{z}$, the
two-qubit reduced density matrix for any two spins in an eigenstate of
$H_{HSG}$ always takes the restricted form

\begin{equation}\label{RDM}
    \rho = \left( \begin{array}{cccc} \alpha & 0 & 0 & 0 \\ 
    0 & \beta & \gamma & 0 \\ 0 & \gamma^{*} & \delta
    & 0 \\ 0 & 0 & 0 & \epsilon \end{array} \right) .
\end{equation}
For such a density matrix, the concurrence is $C(\rho) = 2 \max \left(
0, \vert \gamma \vert - \sqrt{\alpha \epsilon} \right)$.  Concurrence
ranges from zero, for a separable state, to one, for a maximally
entangled state.  The entanglement of formation, another important
entanglement measure, can be calculated directly from the concurrence,
and is monotonically related to it.  The method of calculating the
concurrence for more general density matrices can be found in Wootters
\cite{concurrence}.

In Ref.~\onlinecite{ArnesenBoseVedral} the authors have reported the
absence of thermal and magnetic entanglement in the ferromagnetic
Heisenberg spin chain, based on the analytical solution of the $N=2$
case and numerical studies of $N \leq 10$ chains.  The following
argument shows that the absence of thermal entanglement, at least at
low temperatures, is a general feature of exchange-coupled
ferromagnets, irrespective of the range of the exchange interactions,
the dimensionality of the solid, or the precise boundary conditions
(free or periodic), even in the presence of inhomogeneous coupling
strengths.  Consider the ferromagnetic Heisenberg spin graph
Hamiltonian for $N$ spins: $H_{HSG}=\sum_{i<j} J_{ij} {\mathbf{S}}_{i}
\boldsymbol{\cdot} {\mathbf{S}}_{j}$, where $J_{ij} \leq 0$.  Because
the exchange interaction can be rewritten in terms of a permutation
operator and the identity operator as ${\mathbf{S}}_{i}
\boldsymbol{\cdot} {\mathbf{S}}_{j} = \frac{1}{2} \left(
{\mathcal{P}}_{ij} - \frac{1}{2} {\mathcal{I}} \right)$, the $N+1$
completely symmetric states

\begin{eqnarray}\label{symmetric states}
    \vert \psi_{0} \rangle & = & \vert \emptyset \rangle \\ \nonumber
    \vert \psi_{1} \rangle & = & \frac{1}{\sqrt{N}} \left( \vert 1 \rangle + 
    \ldots + \vert N \rangle \right) \\ \nonumber
    \vert \psi_{2}\rangle  & = & \frac{1}{\sqrt{N(N-1)/2}} \left( \vert 12 
    \rangle + \vert 13 \rangle + 
    \ldots + \vert (N-1) N \rangle \right) \\ \nonumber
    & \vdots & \\  \nonumber
    \vert \psi_{N} \rangle & = & \vert 12 \ldots N \rangle \nonumber
\end{eqnarray}
are eigenstates of $H_{HSG}$ with the common eigenvalue $\frac{1}{4}
\sum_{i<j} J_{ij}$.  These states span the ground subspace of the
model; there are no `accidental' degeneracies (see Appendix).  At
sufficiently low temperature, the thermal density matrix is an
incoherent mixture of these $N+1$ states, the contribution from other
states being negligible.

We choose any two spins in the solid and trace out the rest.  Because
of the linearity of the partial trace, we can perform the trace on
each of the completely symmetric states separately.  The total reduced
density matrix $\rho$ will then be

\begin{equation}\label{def:sumDM}
    \rho = \frac{1}{N+1} \left( \rho_{0} + \ldots + \rho_{N} \right) ,
\end{equation}
where $\rho_{n}$, the reduced density matrix of the $n$th completely
symmetric state, has the form: 

\begin{equation}\label{DM}
    \rho_{n} = \left( \begin{array}{cccc} \alpha_{n} & 0 & 0 & 0 \\ 
    0 & \beta_{n} & \gamma_{n} & 0 \\ 0 & \gamma^{*}_{n} & \delta_{n}
    & 0 \\ 0 & 0 & 0 & \epsilon_{n} \end{array} \right) .
\end{equation}
From combinatoric reasoning we have

\begin{eqnarray}\label{DMentries}
    \alpha_{n} & = & \frac{n(n-1)}{N(N-1)} , \\
    \beta_{n} & = & \gamma_{n} \ = \ \delta_{n} \  = \ \frac{n(N-n)}{N(N-1)} , \\
    \epsilon_{n} & = & \frac{(N-n)(N-n-1)}{N(N-1)} ,
\end{eqnarray}
for the components of the density matrix $\rho_{n}$.

Carrying out the summation we find that the ground state density matrix is

\begin{equation}\label{ferromagneticDM}
    \rho = \left( \begin{array}{cccc} \frac{1}{3} & 0 & 0 & 0 \\ 
    0 & \frac{1}{6} & \frac{1}{6} & 0 \\ 0 & \frac{1}{6} & \frac{1}{6}
    & 0 \\ 0 & 0 & 0 & \frac{1}{3} \end{array} \right) .
\end{equation}
Thus the low temperature density matrix for any two spins assumes a
universal form, irrespective of the strength, geometry, or homogeneity
of the interactions or the dimensionality of the solid.  The
concurrence of this universal density matrix is zero.  This shows
that, at least at low temperatures, no two qubits in an isotropic
ferromagnet share any entanglement.  The correlations present, such as
$\langle S^{x}_{i} S^{x}_{j} \rangle = 1/12$, can be regarded as
purely classical.

The form of the reduced density matrix given in
Eq.~(\ref{ferromagneticDM}) is of course strictly valid only in the
limit of zero temperature.  As the temperature $T$ increases, the
remaining eigenstates of $H_{HSG}$ (those not listed in
Eq.~(\ref{symmetric states})) begin to contribute to the reduced
density matrix $\rho$ describing the state of the two chosen spins. 
As there are a finite number of such states, it can be seen that the
elements $\alpha,\beta,\ldots$, of the reduced density matrix are
continuous functions of $T$.  Since at zero temperature $\vert \gamma
\vert - \sqrt{\alpha\epsilon} = -1/6$, there must be some (possibly
infinite) temperature interval, with $T=0$ as one endpoint and closed
if the interval is finite, in which the concurrence is zero.  Thus the
onset of entanglement in ferromagnets, if it occurs at all, happens as
a phase transition at nonzero temperature.

\begin{figure}
\begin{center}
     \includegraphics{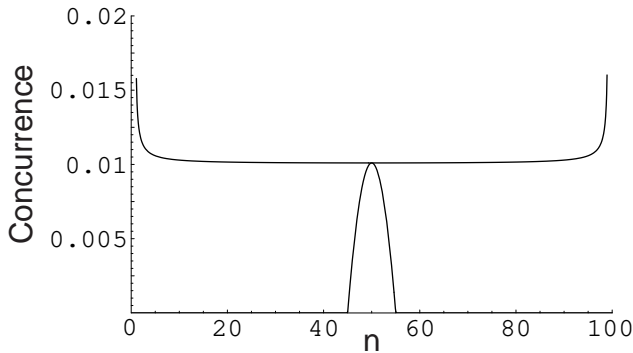}
\end{center}
\caption{Concurrence of the unmixed symmetric state $\rho_{n}$
\emph{(upper curve)}, Eq.~\ref{symmetric state concurrence}, and of
the pairwise mixed symmetric state $\rho'_{n}$ (\emph{lower curve}),
Eq.~\ref{pairwise mixed concurrence}, for $N=100$.}
\label{Figure1}
\end{figure}

\begin{figure}
\begin{center}
     \includegraphics{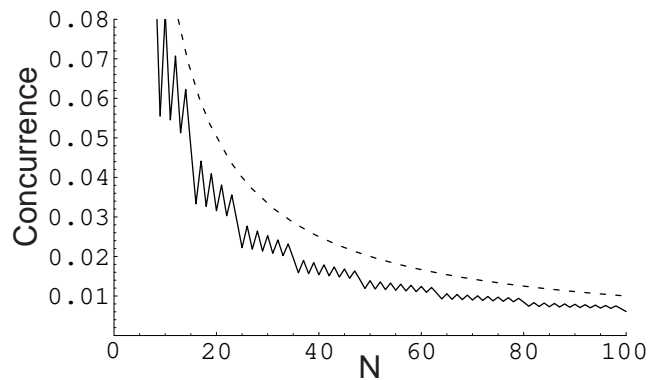}
\end{center}
\caption{Concurrence of the incoherently mixed symmetric states in the
zone $(N - \sqrt{N})/2 < n < (N + \sqrt{N})/2$ as a function of $N$. 
Within this zone pairwise entanglement cancellation fails.  The curve
$1/N$ (\emph{dashed}) is shown for comparison.}
\label{Figure2}
\end{figure}

Note that, individually, each $\rho_{n}$ is an inseparable state,
except for $\rho_{0}$ and $\rho_{N}$, with concurrence

\begin{eqnarray}\label{symmetric state concurrence}
    && C(\rho_{n}) = \frac{2}{N(N-1)}  \times \\
    && \left( n(N-n) - \sqrt{n(n-1)(N-n)(N-n-1)} \right). \nonumber
\end{eqnarray}
It is the incoherent mixing of these inseparable states that washes
out the quantum correlations.  One might anticipate that this
entanglement cancellation occurs pairwise, between states related by
flipping the orientation of each spin, so that $\rho'_{n} \equiv
\left( \rho_{n} + \rho_{N-n} \right) /2$ would be separable; only the
entanglement of $\rho_{N/2}$ when $N$ is even could not be eliminated
in this way.  Interestingly, this is not \emph{quite} correct. 
Rather, there are threshold values of $n$ at $n = (N \pm \sqrt{N})/2$,
below and above which pairwise cancellation occurs, but between which
two-state mixing cannot account for the loss of entanglement. 
Analytically,

\begin{equation}\label{pairwise mixed concurrence}
    C( \rho'_{n} ) = \max \left( 0, \frac{4n(N-n)}{N(N-1)} - 1 \right) ,
\end{equation}
with maximal entanglement of $1/(N-1)$ occuring at $N/2$ ($N$ even) or
of $1/N$ at $(N \pm 1)/2$ ($N$ odd).  Thus it is the \emph{least}
entangled states which are most resistent to pairwise entanglement
cancellation, as illustrated in Fig.~\ref{Figure1}.  Further, the lack
of quantum correlations in the ground subspace cannot be explained by
incoherent mixing of the states within the zone $(N - \sqrt{N})/2 < n
< (N + \sqrt{N})/2$ among themselves \cite{A086520}.  This is already
apparent from consideration of the $N=2$ and $N=4$ cases, where
$\rho_{1}$ and $\rho_{2}$ respectively have no other states in their
zones with which to mix.  But it remains true at $N=6$, where mixing
of $\rho_{2}$, $\rho_{3}$, and $\rho_{4}$ results in an inseparable
state ($C=1/9$).  As shown in Fig.~\ref{Figure2}, the concurrence of
the incoherent mixture of the states within this zone decays as $\sim
1/N$.  The separability of the degenerate ground state therefore
arises from the mixing of this inseparable core with the surrounding
sea of pairwise-separable states.

Whether any isotropic ferromagnet can become entangled at any
temperature remains an open question.  It seems unlikely.  If the
first excited state is nondegenerate, or nearly so, then it will
change the low-temperature density matrix $\rho$ (Eq.~\ref{def:sumDM})
by terms of order $\exp (-\Delta E/kT) /N$, where $\Delta E >0$ is the
energy difference between the ground and first excited state.  Such
changes can be seen to be too small to result in an inseparable state,
by considering the extreme case where the reduced density matrix for
the excited state is a perfectly entangled triplet state.  On the
other hand, if the first excited state is highly degenerate (as in the
Heisenberg spin chain, for example), or if there are many closely
spaced levels just above the first excited state, then the thermal
correction can be of order $\exp (-\Delta E/kT)$, but in this case one
expects that incoherent mixing among the degenerate (or almost
degenerate) excited states will result in a separable state, as
occurred when the degenerate ground states mixed.  In either case,
there appears to be no way to form an inseparable thermal state.

Since the obstacle to thermal entanglement seems to be the highly
degenerate ground state, it is natural to wonder if the application of
an external magnetic field would induce entanglement by relieving the
degeneracy.  To study this question, a simple extension of a
model introduced by Majumdar and Ghosh \cite{MajumdarGhosh}, defined
by the Hamiltonian

\begin{eqnarray}\label{XMGHamiltonian}
     H_{XMG} & = & g_{1} \sum_{i=1}^{N} {\mathbf{S}}_{i}
     \boldsymbol{\cdot} {\mathbf{S}}_{i+1} + g_{2} \sum_{i=1}^{N}
     {\mathbf{S}}_{i} \boldsymbol{\cdot} {\mathbf{S}}_{i+2} \nonumber
     \\ & + & g_{3} \sum_{i=1}^{N} {\mathbf{S}}_{i} \boldsymbol{\cdot}
     {\mathbf{S}}_{i+3} + B \sum_{i=1}^{N} S_{i}^{z} ,
\end{eqnarray}
was studied numerically.  Here $B$ is a homogeneous external magnetic
field directed along the $+z$-axis, and periodic boundary conditions
are assumed, so that ${\mathbf{S}}_{m+N} \equiv {\mathbf{S}}_{m}$. 
The $g_{i}$ are coupling constants, determining the strength of the
interaction of a spin with its $i$th nearest neighbor.  When $g_{3}=0$
the model reduces to the Majumdar-Ghosh spin chain; if $g_{2}=0$ as
well, then it is the Heisenberg spin chain \cite{Bethe}.  Without loss
of generality the Hamiltonian can be rescaled so that $g_{1}=- 1$ for
a ferromagnetic chain.  Concurrences were calculated numerically for
each qubit pair in spin chains of lengths $N=4$ through $8$.  The
coupling constants $g_{2}$ and $g_{3}$ were varied between $-4$ and
$0$, while the temperature and magnetic field were varied between $0$
and $N$.  In no case did any qubit pair exhibit any entanglement. 
Similar numerical studies of other geometries were also carried out,
including open spin chains (i.e. without periodic boundary
conditions), nine spins arrranged as a $3 \times 3$ grid with and
without periodic boundary conditions, eight spins arranged as a cube,
and several irregular geometries.  These calculations also failed to
reveal any entanglement between any qubit pair.  Thus it appears
reasonable to conjecture that ferromagnets cannot exhibit thermal
entanglement at any temperature or applied field strength.

This paper has examined the quantum correlations present in isotropic
ferromagnets.  The ground subspace of an isotropic ferromagnet is
spanned by the completely symmetric states, and it was proven that no
other states intrude.  As a result, the low-temperature two-spin
reduced density matrix assumes a universal, separable form.  Pairwise
entanglement cancellation cannot entirely account for this
separability, as there exists a core of states which remain
inseparable even when mixed incoherently among themselves.  Numerical
evidence was provided supporting the conjecture that isotropic
ferromagnetic spin systems do not exhibit thermal entanglement, even
in the presence of an external magnetic field.

In fact, universality in the entanglement behavior of magnets is even
more general than described here.  Although anisotropic deformations
of the exchange interaction can modify the structure of the ground
subspace in several ways, universal behavior does reappear at low
temperature even in anisotropic models.  This will be discussed in
detail in a subsequent paper.

Acknowledgement: This work was supported in part by the MURI Grant
DAAD19-99-1-0215 and by NSF Grant PHY-0072359.

\appendix*
\section{Appendix}

{\textbf{Theorem:}} The ground subspace of the $N$-site connected
ferromagnetic spin graph, defined by the Hamiltonian $H=\sum_{i<j}
J_{ij} {\mathbf{S}}_{i} {\boldsymbol{\cdot}} {\mathbf{S}}_{j}$ with $J_{ij}
\leq 0$, is spanned by the $N+1$ completely symmetric states given in
Eq.~\ref{symmetric states}; there are no `accidental' degeneracies.

{\textbf{Proof:}} First consider the simpler case where $J_{ij} < 0$,
that is, every spin interacts, though perhaps very weakly, with every
other spin.  The spin operator product ${\mathbf{S}}_{i}
\boldsymbol{\cdot} {\mathbf{S}}_{j}$ is equivalent to $\frac{1}{2}
\left( {\mathcal{P}}_{ij} - \frac{1}{2}{\mathcal{I}} \right)$, where
${\mathcal{P}}_{ij}$ is an operator which permutes spins $i$ and $j$,
and ${\mathcal{I}}$ is the identity operator.  Each completely
symmetric state is an eigenstate of each permutation operator
separately, and so its energy eigenvalue is simply the sum of the
coefficients of the individual permutation operators in the
Hamiltonian: $E_{0} = \langle H \rangle = \frac{1}{2} \sum_{i<j}
J_{ij} \langle {\mathcal{P}}_{ij} - \frac{1}{2}{\mathcal{I}} \rangle =
\frac{1}{4} \sum_{i<j} J_{ij}$.  If a state $\vert \chi \rangle$ is
not completely symmetric, however, there is by definition some
permutation operator which does not leave this state invariant.  In
the special case under consideration, such a permutation has a nonzero
coefficient in the Hamiltonian, say $J_{pq}$.  Now,
${\mathcal{P}}_{ij}^{2} = {\mathcal{I}}$, hence the maximum
expectation value of a permutation operator in a normalized state is
unity, so that $J_{pq} \langle \chi \vert {\mathcal{P}}_{pq} \vert
\chi \rangle > J_{pq}$, since the $J_{ij}$ are negative numbers.  So
$\langle \chi \vert H \vert \chi \rangle = \frac{1}{2} \sum_{i<j}
J_{ij} \langle \chi \vert {\mathcal{P}}_{ij} -
\frac{1}{2}{\mathcal{I}} \vert \chi \rangle > E_{0}$.  Thus the
completely symmetric states span the ground subspace, as claimed.

Now consider the more general case, where some of the coupling
constants $J_{ij}$ may equal zero, but the resulting graph remains
connected.  Suppose once again that a state $\vert \chi \rangle$ is
not symmetric under the interchange of $p$ and $q$.  By the definition
of connectedness, there is now a chain of spins $i_{1}, i_{2}, \ldots,
i_{n}$ such that each of $J_{p i_{1}}, J_{i_{1} i_{2}}, \ldots,
J_{i_{n} q}$ is nonzero.  Then $H^{2n+1}$, the $(2n+1)th$ power of the
Hamiltonian, contains a product of permutation operators
\begin{equation}    
     {\mathcal{P}}_{p i_{1}} \ldots {\mathcal{P}}_{i_{n-1} i_{n}}
     {\mathcal{P}}_{i_{n} q} {\mathcal{P}}_{i_{n-1} i_{n}} \ldots
     {\mathcal{P}}_{i_{1} i_{2}}{\mathcal{P}}_{p i_{1}}
\end{equation}
which is equal to ${\mathcal{P}}_{p q}$ (from the right, the first
$n+1$ permutations put the spin $p$ in the position $q$, while the
remaining put spin $q$ into position $p$ and sort the intermediate
spins of the chain back into their original positions).  The above
argument can now be repeated.  Any completely symmetric state is an
eigenstate of each term of $H^{2n+1}$, while $\vert \chi \rangle$ is
not, and hence the expectation value of $\langle \chi \vert H^{2n+1}
\vert \chi \rangle \neq E_{0}^{2n+1}$.  Therefore $\vert \chi \rangle$
does not lie in the ground subspace.  $\blacksquare$



\end{document}